\documentclass{article}
\usepackage{graphicx}
\usepackage{amssymb}
\usepackage{amsmath}
\usepackage[english]{babel}

\title{
\includegraphics[width=0.35\textwidth]{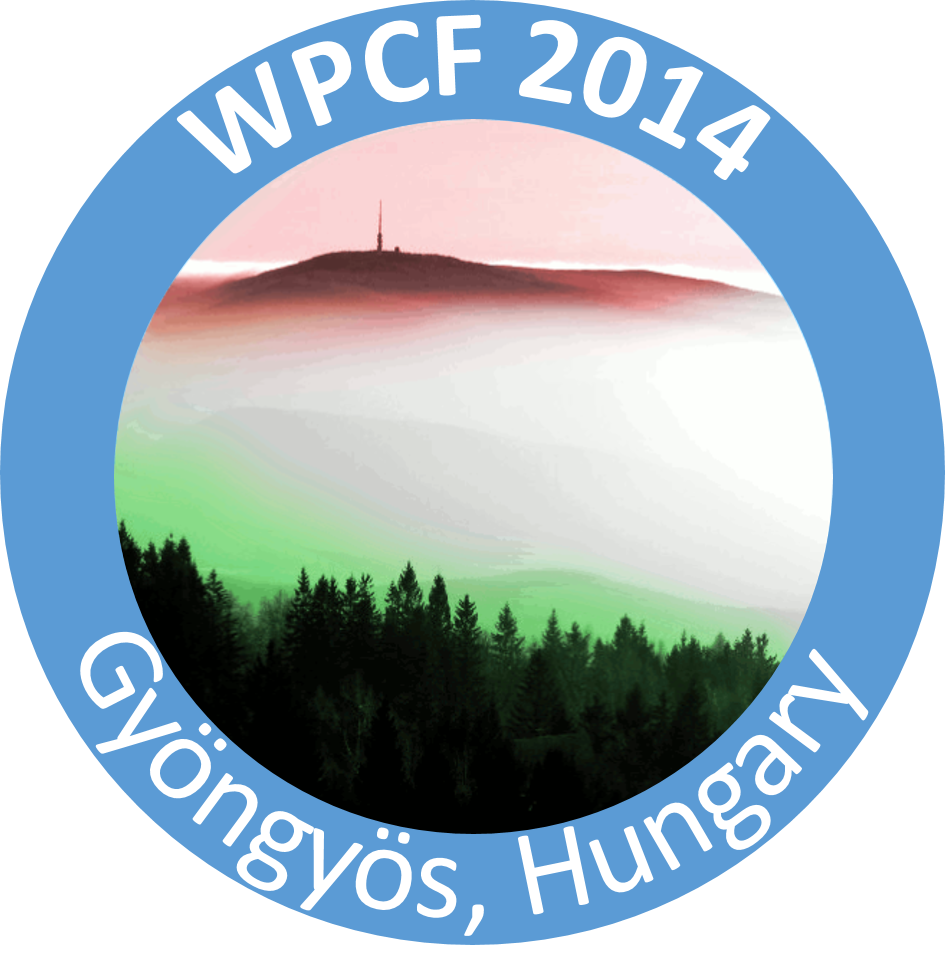}\\[1cm]
Energy dependence of space-time extent of pion source in nuclear
collisions }
\author{{V.A. Okorokov$^1$,}\\[1ex]
$^1$National Research Nuclear University ''MEPhI" \\(Moscow
Engineering Physics Institute), \\ Kashirskoe Shosse 31, 115409
Moscow, Russia\\
}

\begin{document}

\fontfamily{lmss}\selectfont
\maketitle

\begin{abstract}
Energy dependence of space-time parameters of pion emission region
at freeze-out is studied for collisions of various ions and for
all experimentally available energies. The using of femtoscopic
radii scaled on the averaged radius of colliding ions is
suggested. This approach allows the expansion of the set of
interaction types, in particular, on collisions of non-symmetrical
ion beams which can be studied within the framework of common
treatment. There is no sharp changing of femtoscopic parameter
values with increasing of initial energy. Analytic functions
suggested for smooth approximations of energy dependence of
femtoscopic parameters demonstrate reasonable agreement with most
of experimental data at $\sqrt{\smash[b]{s_{\footnotesize{NN}}}}
\geq 5$ GeV. Estimations of some observables are obtained for
energies of the LHC and FCC project.
\end{abstract}

\section{Introduction}\label{sec:1}

At present femtoscopic measurements in particular that based on
Bose\,--\,Einstein correlations are unique experimental method for
the determination of sizes and lifetimes of sources in high energy
and nuclear physics. The study of nucleus-nucleus ($AA$)
collisions in wide energy domain by correlation femtoscopy seems
important for better understanding both the equation of state
(EOS) of strongly interacting matter and general dynamic features
of soft processes. The discussion below is focused on specific
case of femtoscopy, namely, on correlations in pairs of identical
charged pions with small relative momenta -- HBT-interferometry --
in nucleus-nucleus collisions. The general view for
phenomenological parameterization of correlation function (CF) for
two identical particles is discussed in the
\cite{Okorokov-arXiv-1312.4269,Okorokov-arXiv-1409.3925}. Below
the experimental results obtained for $AA$ collisions within the
standard 3d approach are taken into account
\cite{Okorokov-arXiv-1409.3925}. The set of main femtoscopic
observables $\mathcal{G}_{1} \equiv
\{\mathcal{G}_{1}^{i}\}_{i=1}^{4}=\{\lambda,
R_{\mbox{\scriptsize{s}}}, R_{\mbox{\scriptsize{o}}},
R_{\mbox{\scriptsize{l}}}\}$ is under consideration as well as the
set of some important additional observables which can be
calculated with help of HBT radii $\mathcal{G}_{2} \equiv
\{\mathcal{G}_{2}^{j}\}_{j=1}^{3}=\{R_{\mbox{\scriptsize{o}}}/
R_{\mbox{\scriptsize{s}}}, \delta, V\}$. Here $\delta=
R_{\mbox{\scriptsize{o}}}^{2}-R_{\mbox{\scriptsize{s}}}^{2}$,
$V=(2\pi)^{3/2}R_{\mbox{\scriptsize{s}}}^{2}R_{\mbox{\scriptsize{l}}}$
is the volume of source at freeze-out. The set of parameters
$\mathcal{G}_{1}$ characterizes the correlation strength and
source's 4-dimensional geometry at freeze-out stage completely.
The most central collisions are usually used for study the
space-time characteristics of final-state matter, in particular,
for discussion of global energy dependence of femtoscopic
observables. Therefore scaled parameters $\mathcal{G}_{1}^{i}$,
$i=2-4$, $\delta$ and $\mathcal{G}_{2}^{3}$ are calculated as
follows \cite{Okorokov-arXiv-1312.4269,Okorokov-arXiv-1409.3925}:
\begin{equation}
R_{i}^{n}=R_{i}/R_{\mbox{\scriptsize{A}}},~
i=\mbox{s,o,l};~~~\delta^{n}=\delta/R_{\mbox{\scriptsize{A}}}^{2};~~~V^{n}=V/V_{\mbox{\scriptsize{A}}}.\label{eq:2.8}
\end{equation}
Here $R_{\mbox{\scriptsize{A}}}=R_{0}A^{1/3},
V_{\mbox{\scriptsize{A}}}=4\pi R^{3}_{\mbox{\scriptsize{A}}}/3$ is
radius and volume of spherically-symmetric nucleus, $R_{0}=(1.25
\pm 0.05)$ fm \cite{Valentin-book-1982,Mukhin-book-1983}. The
change $R_{\mbox{\scriptsize{A}}} \to \langle
R_{\mbox{\scriptsize{A}}}\rangle=0.5(R_{\mbox{\scriptsize{A}}_{1}}+R_{\mbox{\scriptsize{A}}_{2}})$
is made in the relation (\ref{eq:2.8}) in the case of
non-symmetric nuclear collisions
\cite{Okorokov-arXiv-1312.4269,Okorokov-arXiv-1409.3925}. In
general case the scale factor in (\ref{eq:2.8}) should takes into
account the centrality of nucleus-nucleus collisions. The
normalization procedure suggested in
\cite{Okorokov-arXiv-1312.4269} allows the consideration of all
available data for nucleus-nucleus collisions
\cite{Okorokov-arXiv-1409.3925}. As development of previous
analyses \cite{Okorokov-arXiv-1312.4269,Okorokov-arXiv-1409.3925}
the proton-proton ($pp$) results at high energies
\cite{STAR-PRC-83-064905-2011,ALICE-PRD-84-112004-2011} are also
considered here with replacing $R_{\mbox{\scriptsize{A}}} \to
R_{\mbox{\scriptsize{p}}}$ in (\ref{eq:2.8}).

\section{Energy dependence of space-time extent of pion source}\label{sec:2}
\begin{figure*}[h!]
\begin{center}
\includegraphics[width=12.0cm,height=12.0cm]{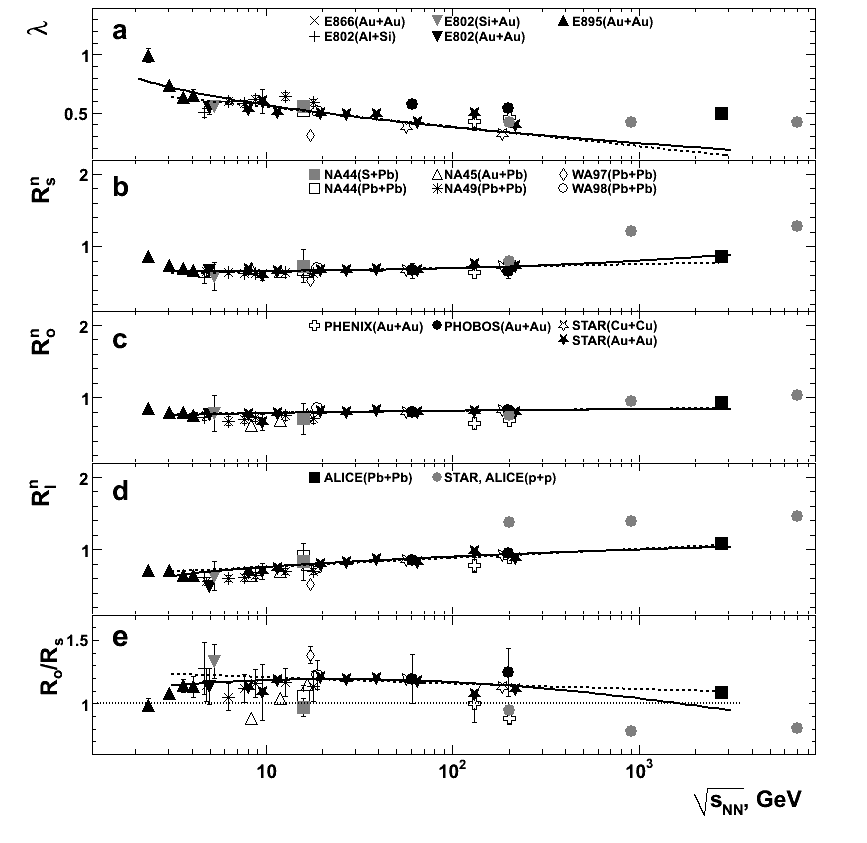}
\end{center}
\vspace*{-0.8cm} \caption{Energy dependence of $\lambda$ parameter
(a), scaled HBT-radii (b -- d) and ratio
$R_{\mbox{\scriptsize{o}}}/R_{\mbox{\scriptsize{s}}}$ (e) in
various collisions. Experimental data are from
\cite{Okorokov-arXiv-1409.3925,STAR-PRC-83-064905-2011,ALICE-PRD-84-112004-2011}.
Statistical errors are shown (for NA44 -- total uncertainties).
The solid lines (a -- d) correspond to the fits by function
(\ref{eq:Fit-1}) and dashed lines -- to the fits by specific case
of (\ref{eq:Fit-1}) at fixed $a_{3}=1.0$. Smooth solid and dashed
curves at (e) correspond to the ratio
$R_{\mbox{\scriptsize{o}}}/R_{\mbox{\scriptsize{s}}}$ calculated
from the fit results for $R^{n}_{\mbox{\scriptsize{s}}}$ and
$R^{n}_{\mbox{\scriptsize{o}}}$ in $AA$, dotted line is the level
$R_{\mbox{\scriptsize{o}}}/R_{\mbox{\scriptsize{s}}}=1$.}
\label{fig:4}
\end{figure*}
\begin{table*}[h!]
\caption{Values of fit parameters for $AA$ data with statistical
errors} \label{tab:3}
\begin{center}
\begin{tabular}{lcccc}
\hline \multicolumn{1}{l}{HBT} & \multicolumn{4}{c}{Fit
parameter}\rule{0pt}{10pt}\\
\cline{2-5}
parameter & $a_{1}$ & $a_{2}$ & $a_{3}$ & $\chi^{2}/\mbox{n.d.f.}$\rule{0pt}{10pt}\\
\hline
$\lambda$                       & $1.21 \pm 0.09$   & $-0.30 \pm 0.04$           & $0.38 \pm 0.04$   & $3656/29$ \rule{0pt}{10pt}\\
                                & $0.717 \pm 0.003$ & $-0.051 \pm 0.001$         & $1.0$ (fixed)     & $3786/23$ \rule{0pt}{10pt}\\
$R^{n}_{\mbox{\scriptsize{s}}}$ & $0.656 \pm 0.002$ & $(6 \pm 3) \times 10^{-5}$ & $3.11 \pm 0.19$   & $195/25$  \rule{0pt}{10pt}\\
                                & $0.599 \pm 0.003$ & $0.019 \pm 0.001$          & $1.0$ (fixed)     & $280/26$  \rule{0pt}{10pt}\\
$R^{n}_{\mbox{\scriptsize{o}}}$ & $0.10 \pm 0.02$   & $6.3 \pm 1.7$              & $0.068 \pm 0.006$ & $402/25$  \rule{0pt}{10pt}\\
                                & $0.758 \pm 0.004$ & $0.008 \pm 0.001$          & $1.0$ (fixed)     & $415/26$  \rule{0pt}{10pt}\\
$R^{n}_{\mbox{\scriptsize{l}}}$ & $0.022 \pm 0.002$ & $23 \pm 3$                 & $0.258 \pm 0.005$ & $502/25$  \rule{0pt}{10pt}\\
                                & $0.634 \pm 0.004$ & $0.043 \pm 0.001$          & $1.0$ (fixed)     & $615/26$  \rule{0pt}{10pt}\\
\hline
\end{tabular}
\end{center}
\end{table*}

Detail study for (quasi)symmetric heavy ion collisions
\cite{Okorokov-arXiv-1312.4269,Okorokov-arXiv-1409.3925}
demonstrates that the fit function ($\varepsilon \equiv
s_{\footnotesize{NN}}/s_{0}$, $s_{0}=1$ GeV$^{2}$)
\begin{equation}
f(\sqrt{\smash[b]{s_{\footnotesize{NN}}}}) = a_{1}\left[1 +
a_{2}(\ln\varepsilon)^{a_{3}}\right] \label{eq:Fit-1}
\end{equation}
agrees reasonably with experimental dependence
$\mathcal{G}_{1}^{i}(\sqrt{\smash[b]{s_{\footnotesize{NN}}}})$,
$i=1-4$ at any collision energy for $\lambda$ and at
$\sqrt{\smash[b]{s_{\footnotesize{NN}}}} \geq 5$ GeV for HBT
radii. Fig.\,\ref{fig:4} shows the energy dependence of $\lambda$
(a), scaled HBT-radii (b -- d) and
$R_{\mbox{\scriptsize{o}}}/R_{\mbox{\scriptsize{s}}}$ ratio (e)
for both the symmetric and non-symmetric collisions of various
nuclei. Fits of experimental dependencies for AA interactions are
made by (\ref{eq:Fit-1}) in the same energy domains as well as for
(quasi)symmetric heavy ion collisions. The numerical values of fit
parameters are presented in Table\,\ref{tab:3}, fit curves are
shown in Fig.\,\ref{fig:4} by solid lines for (\ref{eq:Fit-1}) and
by dashed lines for specific case of fit function at $a_{3}=1.0$
with taking into account statistical errors. There is dramatic
growth of $\chi^{2}/\mbox{n.d.f.}$ values for fits of $\lambda$
data (Fig.\,\ref{fig:4}a) despite of qualitative agreement between
smooth approximations and experimental $\lambda$ values for range
$10 \lesssim \sqrt{\smash[b]{s_{\footnotesize{NN}}}} \lesssim 200$
GeV. The fit by (\ref{eq:Fit-1}) underestimates the $\lambda$
value at the LHC energy
$\sqrt{\smash[b]{s_{\footnotesize{NN}}}}=2.76$ TeV significantly.
The $\lambda$ values for asymmetric nucleus-nucleus collisions at
intermediate energies $\sqrt{\smash[b]{s_{\footnotesize{NN}}}}
\lesssim 20$ GeV agree well with values of $\lambda$ in symmetric
heavy ion collisions at close energies. On the other hand the
$\lambda$ for $\mbox{Cu+Cu}$ collisions is smaller systematically
than $\lambda$ in $\mbox{Au+Au}$ collisions in energy range
$\sqrt{\smash[b]{s_{\footnotesize{NN}}}}=62-200$ GeV
(Fig.\,\ref{fig:4}a). New experimental data are important for
verification of the suggestion of separate dependencies
$\lambda(\sqrt{\smash[b]{s_{\footnotesize{NN}}}})$ for moderate
and heavy ion collisions. Also the development of some approach is
required in order to account for type of colliding beams in the
case of $\lambda$ parameter and improve quality of approximation.
Smooth curves for normalized HBT radii and ratio
$R_{\mbox{\scriptsize{o}}}/R_{\mbox{\scriptsize{s}}}$ are in
reasonable agreement with experimental dependencies in fitted
domain of collision energies
$\sqrt{\smash[b]{s_{\footnotesize{NN}}}} \geq 5$ GeV
(Figs.\,\ref{fig:4}b -- e). Dramatic improvement of the fit
qualities for scaled HBT radii at transition from the data sample
with statistical errors to the data sample with total errors is
dominated mostly by the uncertainty in $r_{0}$ leads to additional
errors due to scaling (\ref{eq:2.8}). The scaled HBT-radii in $pp$
are larger significantly than those in $AA$ collisions at close
energies. Because feature of Regge theory \cite{Collins-book-1977}
the following relation is suggested to take into account the
expanding of proton with energy:
$R_{\mbox{\scriptsize{p}}}=r_{0}(1+k\sqrt{\alpha'_{\cal{P}}\ln\varepsilon})$,
where $r_{0}=(0.877 \pm 0.005)$ fm is the proton's charge radius
\cite{PDG-PRD-86-010001-2012}, parameter $\alpha'_{\cal{P}}
\propto \ln\varepsilon$ because of diffraction cone shrinkage
speeds up with collision energy in elastic $pp$ scattering
\cite{Okorokov-arXiv-1501.01142}. The $k$ is defined from the
boundary condition $R_{\mbox{\scriptsize{p}}} \to 1/m_{\pi}$ at
$\varepsilon \to \infty$ with choice of appropriate asymptotic
energy $\sqrt{s_{\footnotesize{NN}}^{\footnotesize{a}}}$. The
detail study demonstrates that the increasing of
$\sqrt{s_{\footnotesize{NN}}^{\footnotesize{a}}}$ from 6 PeV
\cite{Bourrely-arXiv-1202.3611} to $10^{3}$ PeV influences weakly
on $R^{n}_{\mbox{\scriptsize{i}}}$, $i=\mbox{s}, \mbox{o},
\mbox{l}$ in $pp$ collisions and calculations are made for the
first case. The normalized transverse radii agree in both the $pp$
and the $AA$ collisions (Figs.\,\ref{fig:4}b, c) at
$\sqrt{\smash[b]{s_{\footnotesize{NN}}}}=200$ GeV with excess of
$R^{n}_{\mbox{\scriptsize{s}}}$ in $pp$ with respect to the $AA$
in TeV-region. The $R^{n}_{\mbox{\scriptsize{l}}}$ in $pp$ is
larger than that for $AA$ in domain
$\sqrt{\smash[b]{s_{\footnotesize{NN}}}} \geq 200$ GeV. It should
be stressed that the additional study is important, at least, for
choice of $R_{\mbox{\scriptsize{p}}}(\varepsilon)$.

\begin{figure*}[h!]
\begin{center}
\includegraphics[width=8.0cm,height=8.0cm]{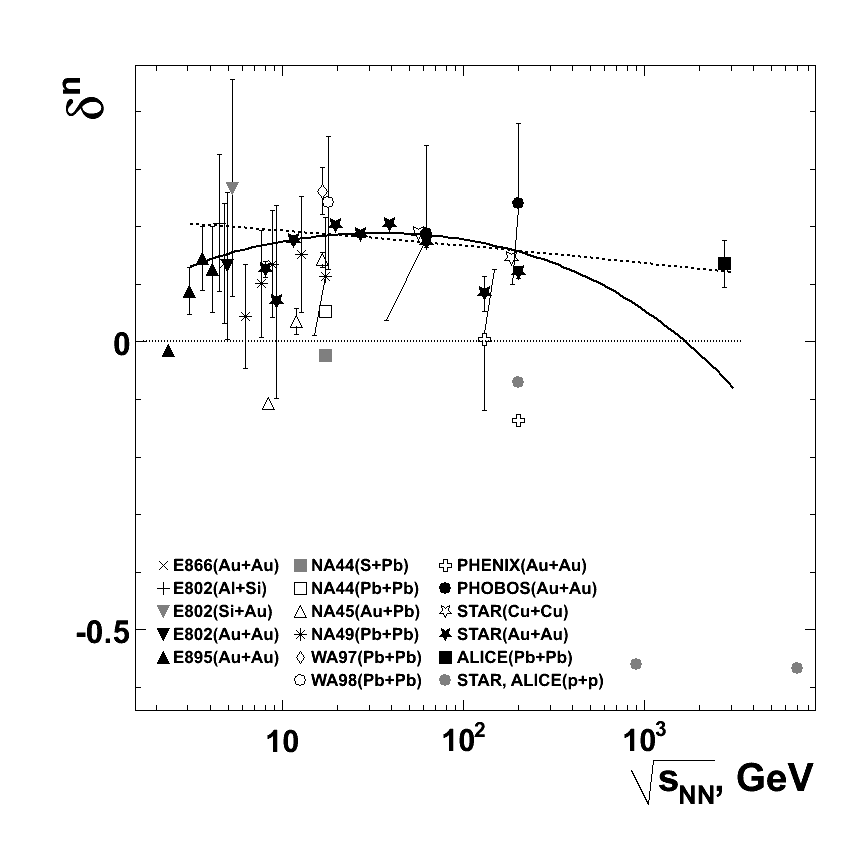}
\end{center}
\vspace*{-0.8cm} \caption{Dependence
$\delta^{n}(\sqrt{\smash[b]{s_{\footnotesize{NN}}}})$ in various
collisions. Experimental data are from
\cite{Okorokov-arXiv-1409.3925,STAR-PRC-83-064905-2011,ALICE-PRD-84-112004-2011}.
Error bars are only statistical (for NA44 -- total uncertainties).
Dotted line is the level $\delta^{n}=0$. Smooth curves are derived
from (\ref{eq:2.8}) and the fit results for
$R^{n}_{\mbox{\scriptsize{s}}}$, $R^{n}_{\mbox{\scriptsize{o}}}$
in $AA$. The solid line corresponds to the fits of normalized HBT
radii by function (\ref{eq:Fit-1}) and dashed line -- to the fits
by specific case $R^{n}_{i} \propto \ln\varepsilon$, $i=\mbox{s},
\mbox{o}$.} \label{fig:5}
\end{figure*}

\begin{figure*}[h!]
\begin{center}
\includegraphics[width=8.0cm,height=8.0cm]{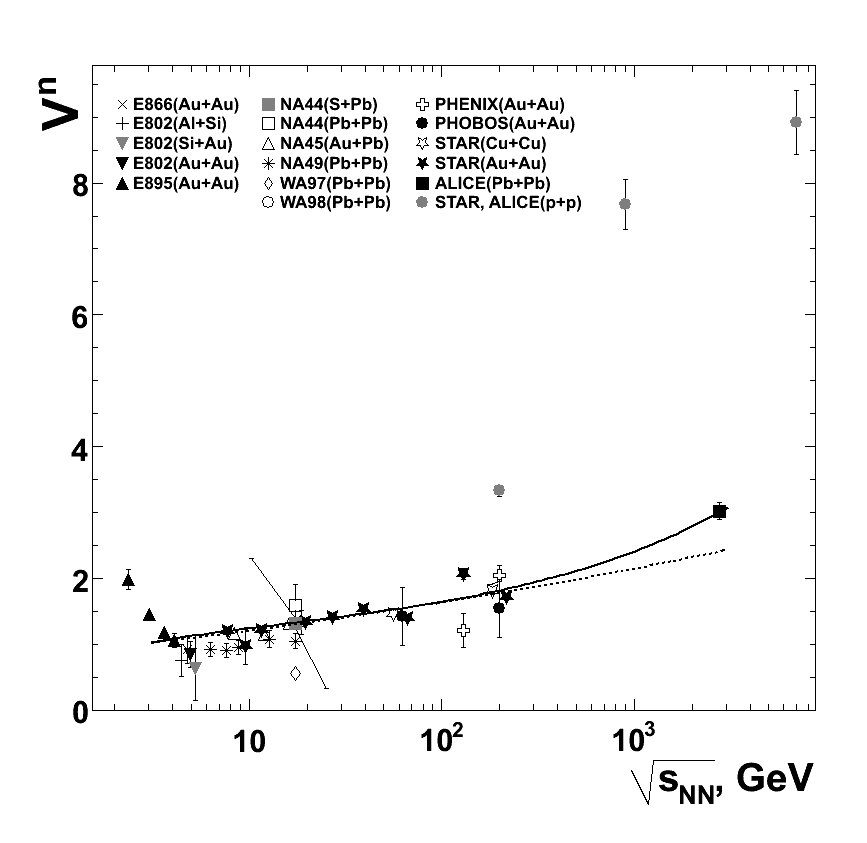}
\end{center}
\vspace*{-0.8cm} \caption{Energy dependence of $V^{n}$ in various
collisions. Experimental data are from
\cite{Okorokov-arXiv-1409.3925,STAR-PRC-83-064905-2011,ALICE-PRD-84-112004-2011}.
Error bars are only statistical (for NA44 -- total uncertainties).
Smooth curves are derived from (\ref{eq:2.8}) and the fit results
for $R^{n}_{\mbox{\scriptsize{s}}}$,
$R^{n}_{\mbox{\scriptsize{l}}}$ in $AA$. The solid line
corresponds to the fits of normalized HBT radii by function
(\ref{eq:Fit-1}) and dashed line -- to the fits by specific case
$R^{n}_{i} \propto \ln\varepsilon$, $i=\mbox{s}, \mbox{l}$.}
\label{fig:6}
\end{figure*}

The corresponding dependencies for $\delta^{n}$ and $V^{n}$ are
demonstrated in Fig.\,\ref{fig:5} and Fig.\,\ref{fig:6},
respectively. As well as in
\cite{Okorokov-arXiv-1312.4269,Okorokov-arXiv-1409.3925} results
for $\pi^{+}\pi^{+}$ pairs are shown in Figs.\,\ref{fig:4} --
\ref{fig:6} also because fem\-to\-sco\-py parameters from the set
$\mathcal{G}_{1}$ depend on sign of electrical charge of secondary
pions weakly. The relation $R_{\mbox{\scriptsize{o}}} <
R_{\mbox{\scriptsize{s}}}$ is observed for $\approx 11\%$ of
points in Fig.\,\ref{fig:5}. Detail discussion for points with
$\delta < 0$ is in the \cite{Okorokov-arXiv-1409.3925}. The
dependence $\delta^{n}(\sqrt{\smash[b]{s_{\footnotesize{NN}}}})$
is almost flat within large error bars in all energy domain under
consideration. Taking into account the STAR high-statistics
results \cite{STAR-arXiv-1403.4972} only one can see the
indication on change of behavior of
$\delta^{n}(\sqrt{\smash[b]{s_{\footnotesize{NN}}}})$ inside the
range of collision energy
$\sqrt{\smash[b]{s_{\footnotesize{NN}}}}=11.5-19.6$ GeV. This
observation is in agreement with features of behavior of emission
duration $(\Delta\tau)$ dependence on
$\sqrt{\smash[b]{s_{\footnotesize{NN}}}}$ discussed in
\cite{Okorokov-arXiv-1409.3925}. The estimation of energy range
agrees well with results of several studies
in the framework of the phase-I of the beam energy scan (BES)
program at RHIC which indicate on the transition from dominance of
quark-gluon degrees of freedom to hadronic matter at
$\sqrt{\smash[b]{s_{\footnotesize{NN}}}} \lesssim 19.6$ GeV. But
future precise measurements are crucially important for extraction
of more definite physics conclusions. Smooth solid and dashed
curves shown in Fig.\,\ref{fig:5} are calculated for $\delta^{n}$
from the fit results for $R^{n}_{\mbox{\scriptsize{s}}}$ and
$R^{n}_{\mbox{\scriptsize{o}}}$ (Table\,\ref{tab:3}). The
calculation based on the fit function (\ref{eq:Fit-1}) at free
$a_{3}$ agrees reasonably with experimental points at
$\sqrt{\smash[b]{s_{\footnotesize{NN}}}} \leq 200$ GeV but
underestimates $\delta^{n}$ in TeV-region significantly. The large
errors in Fig.\,\ref{fig:6} for strongly asymmetric $AA$
collisions is dominated by large difference of radii of colliding
moderate and heavy nuclei and corresponding large uncertainty for
$\langle R_{\mbox{\scriptsize{A}}}\rangle$. Smooth solid and
dashed curves shown in Fig.\,\ref{fig:6} are calculated for
$V^{n}$ from it's definition (\ref{eq:2.8}) and the fit results
for $R^{n}_{\mbox{\scriptsize{s}}}$,
$R^{n}_{\mbox{\scriptsize{l}}}$ (Table\,\ref{tab:3}). The fit
results for normalized HBT radii obtained with general function
(\ref{eq:Fit-1}) lead to very good agreement between smooth curve
and experimental data in TeV-region in contrast with the curve
obtained from corresponding fit results for (\ref{eq:Fit-1}) at
$a_{3}=1.0$. There is significant difference between $pp$ and $AA$
collisions for $\delta^{n}$ in TeV-region (Fig.\,\ref{fig:5}) and
for $V^{n}$ at $\sqrt{\smash[b]{s_{\footnotesize{NN}}}} \geq 200$
GeV (Fig.\,\ref{fig:6}).

Estimations for $\lambda$,
$R_{\mbox{\scriptsize{o}}}/R_{\mbox{\scriptsize{s}}}$, and
normalized femtoscopic parameters at the LHC and the FCC energies
are shown in Table\,\ref{tab:4} for fits of various $AA$
collisions with inclusion of statistical errors, the second line
for each collision energy corresponds to the using of the specific
case of (\ref{eq:Fit-1}) at $a_{3}=1.0$. All the smooth
approximations discussed above predict amplification of coherent
pion emission with significant decreasing of $\lambda$.
Uncertainties are large for estimations obtained on the basis of
results of fits by function (\ref{eq:Fit-1}) at free $a_{3}$. Thus
values of femtoscopic observables in Table\,\ref{tab:4} are equal
within errors for general and specific case of (\ref{eq:Fit-1}) at
$\sqrt{\smash[b]{s_{\footnotesize{NN}}}}=5.52$ TeV (LHC) and
$\sqrt{\smash[b]{s_{\footnotesize{NN}}}}=39.0$ TeV (FCC).

\begin{table*}[h!]
\caption{Estimations for observables based on fit results}
\label{tab:4}
\begin{center}
\begin{tabular}{lcccc}
\hline
\multicolumn{1}{l}{$\sqrt{\smash[b]{s_{\footnotesize{NN}}}}$,} &
\multicolumn{4}{c}{HBT parameter for $AA$}\rule{0pt}{10pt}\\
\cline{2-5}
TeV & $\lambda$ & $R^{n}_{\mbox{\scriptsize{s}}}$ & $R^{n}_{\mbox{\scriptsize{o}}}$ & $R^{n}_{\mbox{\scriptsize{l}}}$ \rule{0pt}{10pt}\\
\hline
5.52 & $0.16 \pm 0.19$   & $0.9 \pm 0.2$     & $0.8 \pm 0.3$     & $1.06 \pm 0.16$   \rule{0pt}{10pt}\\
     & $0.091 \pm 0.004$ & $0.792 \pm 0.009$ & $0.860 \pm 0.010$ & $1.099 \pm 0.013$ \rule{0pt}{10pt}\\
39.0 & $0.07 \pm 0.21$   & $1.2 \pm 0.4$     & $0.9 \pm 0.3$     & $1.11 \pm 0.16$   \rule{0pt}{10pt}\\
     & --                & $0.836 \pm 0.011$ & $0.883 \pm 0.012$ & $1.205 \pm 0.015$ \rule{0pt}{10pt}\\
\hline
     & $R_{\mbox{\scriptsize{o}}}/R_{\mbox{\scriptsize{s}}}$ & $\delta^{n}$ & $V^{n}$ & \rule{0pt}{10pt}\\
\hline
5.52 & $0.9 \pm 0.4$     & $-0.2 \pm 0.6$  & $3.5 \pm 1.6$ \rule{0pt}{10pt}\\
     & $1.086 \pm 0.018$ & $0.11 \pm 0.02$ & $2.59 \pm 0.07$ \rule{0pt}{10pt}\\
39.0 & $0.7 \pm 0.3$     & $-0.7 \pm 1.1$  & $6 \pm 4$      \rule{0pt}{10pt}\\
     & $1.06 \pm 0.02$   & $0.08 \pm 0.03$ & $3.17 \pm 0.09$ \rule{0pt}{10pt}\\
\hline
\end{tabular}
\end{center}
\end{table*}

The energy dependencies for sets $\mathcal{G}_{m}$, $m=1, 2$ of
femtoscopic parameters with taking into account the scaling
relation (\ref{eq:2.8}) demonstrate the reasonable agreement
between values of parameters obtained for interactions of various
ions (Figs.\,\ref{fig:4} -- \ref{fig:6}). The observation confirms
the suggestion \cite{Okorokov-arXiv-1312.4269} that normalized
femtoscopic parameters allow us to unite the study both the
symmetric and the asymmetric $AA$ collisions within the framework
of united approach. This qualitative suggestion is confirmed
indirectly by recent study of two-pion correlations in the
collisions of the lightest nucleus ($\mbox{d}$) with heavy ion
($\mbox{Au}$) at RHIC. Estimations of space-time extent of the
pion emission source in $\mbox{d+Au}$ collisions at top RHIC
energy \cite{PHENIX-arXiv-1404.5291} in dependence on kinematic
observables show similar patterns with corresponding dependencies
in $\mbox{Au+Au}$ collisions and indicate on similarity in
expansion dynamics in collisions of various systems ($\mbox{d+Au}$
and $\mbox{Au+Au}$ at RHIC, $\mbox{p+Pb}$ and $\mbox{Pb+Pb}$ at
LHC). The scaling results for some radii indicate that
hydrodynamic-like collective expansion is driven by final-state
rescattering effects \cite{PHENIX-arXiv-1404.5291}. On the other
hand the normalized femtoscopic parameters allow us to get the
common kinematic dependencies only without any additional
information about possible general dynamic features in different
collisions. Thus the hypothesis discussed above is qualitative
only. The future quantitative theoretical and phenomenological
studies are essential for verification of general features of soft
stage dynamics for different collisions at high energies.

\section{Summary}\label{sec:3}

The main results of present study are the following.

Energy dependence is investigated for range of all experimentally
available initial energies and for estimations of the main
femtoscopic parameters from set the $\mathcal{G}_{1}$ ($\lambda$
and radii) derived in the framework of Gauss approach as well as
for the set of important additional observables $\mathcal{G}_{2}$
contains ratio of transverse radii, $\delta$ and HBT volume. There
is no dramatic change of femtoscopic parameter values in $AA$ with
increasing of $\sqrt{\smash[b]{s_{\footnotesize{NN}}}}$ in domain
of collision energies $\sqrt{\smash[b]{s_{\footnotesize{NN}}}}
\geq 5$ GeV. The energy dependence is almost flat for the
$\delta^{n}$ in nucleus-nucleus collisions within large error
bars. The indication on possible curve knee at
$\sqrt{\smash[b]{s_{\footnotesize{NN}}}} \sim 10-20$ GeV obtained
in the STAR high-statistics data agree with other results in the
framework of the phase-I of the BES program at RHIC. But
additional precise measurements are crucially important at various
$\sqrt{\smash[b]{s_{\footnotesize{NN}}}}$ in order to confirm this
feature in energy dependence of additional femtoscopic parameters.
The normalized some HBT radii and source volume in $pp$ are larger
significantly than those in $AA$ collisions especially in
TeV-region. The fit curves demonstrate qualitative agreement with
experimental $AA$ data for $\lambda$ at all available collision
energies and for normalized HBT radii in energy domain
$\sqrt{\smash[b]{s_{\footnotesize{NN}}}} \geq 5$ GeV. Smooth
curves calculated for energy dependence of parameters from the set
$\mathcal{G}_{2}$ agree reasonably with corresponding experimental
$AA$ data in the most cases. Estimations of femtoscopic
observables are obtained on the basis of the fit results for
energies of the LHC and the FCC project. For multi-TeV energy
domain the emission region of pions in nucleus-nucleus collisions
will be characterized by decreased correlation strength, linear
sizes about $8.5 - 9.5$ fm in longitudinal direction and $7 - 8$
fm in transverse plane, volume of about $10^{4}$ fm$^{3}$.

\end{document}